\documentclass[pra,twocolumn,showpacs,preprintnumbers,superscriptaddress]
{revtex4}

\usepackage{times}
\usepackage{bm}
\usepackage{float}
\usepackage{graphicx}
\usepackage{amsbsy}
\usepackage{amsmath}
\usepackage{amsfonts}
\usepackage{amsthm}
\begin{document}
	
	\theoremstyle{plain}
	\newtheorem{theorem}{Theorem}
	\newtheorem{lemma}[theorem]{Lemma}
	\newtheorem{corollary}[theorem]{Corollary}
	\newtheorem{proposition}[theorem]{Proposition}
	\newtheorem{conjecture}[theorem]{Conjecture}
	
	\theoremstyle{definition}
	\newtheorem{definition}[theorem]{Definition}
	
	\theoremstyle{remark}
	\newtheorem*{remark}{Remark}
	\newtheorem{example}{Example}
	\title{Structured Negativity: A physically realizable measure of entanglement based on structural physical approximation}
	\author{Anu Kumari, Satyabrata Adhikari}
	\email{mkumari_phd2k18@dtu.ac.in, satyabrata@dtu.ac.in} \affiliation{Delhi Technological
		University, Delhi-110042, Delhi, India}
	
	\begin{abstract}
		Quantification of entanglement is one of the most important problem in quantum information theory. In this work, we will study this problem by defining a physically realizable measure of entanglement for any arbitrary dimensional bipartite system $\rho$, which we named as structured negativity $(N_S(\rho))$. We have shown that the introduced measure satisfies the properties of a valid entanglement monotone. We also have established an inequality that relate negativity and the structured negativity. For $d\otimes d$ dimensional state, we conjecture from the result obtained in this work that negativity coincide with the structured negativity when the number of negative eigenvalues of the partially transposed matrix is equal to $\frac{d(d-1)}{2}$. Moreover, we proved that the structured negativity not only implementable in the laboratory but also a better measure of entanglement in comparison to negativity. In few cases, we obtain that structure negativity gives better result than the lower bound of the concurrence obtained by Albeverio [Phys. Rev. Lett. \textbf{95}, 040504 (2005)].

	\end{abstract}
	 \maketitle

	\section{Introduction}
	\noindent Quantum entanglement \cite{horodeckirev} has been considered as most important non-classical feature of quantum information theory. We may realize its usefulness when we think of quantum networks which is based on non-local feature of the entanglement that may provide the basis of some speculated applications such as large scale quantum computation and distributed quantum computing \cite{leent}.
	
	Quantification of entanglement is one of the crucial task in quantum information theory. The importance of this problem can be understood if we consider a simple instance in which we study the relationship between the amount of entanglement present in the shared arbitrary dimensional bipartite resource state and the fidelity of teleportation \cite{sazim}. The quantification problem has already been studied for two qubit system, bipartite higher dimensional and multi-qubit system but still there exist few problems in higher dimensional bipartite mixed system that need to be addressed. There exist various entanglement measures such as concurrence \cite{wootters,wootters1,wootters2}, negativity\cite{vidal}, relative entropy of entanglement \cite{plenio}, geometric measure of entanglement\cite{wei} that can quantify the amount of entanglement in a two-qubit as well as higher dimensional bipartite pure and mixed state. Now, the question arises that whether the entanglement measures existed in the literature can quantify the amount of entanglement for any arbitrary dimensional bipartite system practically? In case of two qubit state, entanglement of formation \cite{wootters} can be measured without prior state reconstruction \cite{horodecki1}. Also, it has been shown that a single observable is not sufficient to determine the entanglement of a given unknown pure two-qubit state \cite{sancho} but neverthless, the amount of entanglement in a pure two-qubit state can be determined experimentally with a minimum of two-copies of the state \cite{walborn}.\\
	The situation will become more complex, when we will consider the problem of quantification of entanglement for higher dimensional bipartite system. For higher dimensional bipartite pure state, there exist some measures of entanglement such as generalized concurrence \cite{rungta2}, negativity\cite{lee}, geometric measure of entanglement that may quantify the amount of entanglement in the given pure state but on the contrary, we have handful of measures of entanglement which work for higher dimensional bipartite mixed state. This is due to the fact that, till today we don't have any closed formula for the concurrence of higher dimensional bipartite mixed state. Secondly, an easily computable measure of entanglement, namely, negativity which may be used to quantify the amount of entanglement in higher dimensional bipartite pure as well as mixed states but the problem with this measure is that it depends on the negative eigenvalues of the non-physical partial transposition operation. Thus, negativity does not correspond to a completely positive map and hence, difficult to impliment it in the laboratory. Recently, generalized geometric measure of entanglement has been defined for multipartite mixed states \cite{sen} but it is not yet clear whether it can be realizable quantity in experiment or not. Another way of quantification of entanglement is by using witness operators \cite{brandao, guhne2} that can be employed for any arbitrary $d_1 \otimes d_2$ dimensional system. Although witness operator is physically realizable but in general it is not easy to construct witness operator for the detection of entangled state.\\
	Horodecki \cite{horodecki1} have proposed a protocol to directly measure the concurrence of $2\otimes 2$  system using four moments only but to calculate the expectation of these four moments, the method needs atmost 20 copies of the state. Their scheme works using structural physical approximation (SPA) of the partial transposition map. This method is efficient in comparison to quantum state tomography with respect to the estimation of state parameters but on the other hand, this method will show it's inefficiency if we compare the number of copies required in the above mentioned methods. This motivate us to define a new measure of entanglement using structural physical approximation of partial transposition.\\
	In this work, we define a physically realizable measure of entanglement which is based on the minimum eigenvalue of the structural physical approximation of partial transposition (SPA-PT) operation. The concept of SPA has been introduced in \cite{horodecki3}, which approximate a non-physical operation to a physical operation. This implies that the SPA map transform a positive but not completely positve maps to a completely positive map. A lot of progress has been made in both theoretical and experimental aspects of SPA\cite{bae,lewenstein3,lewenstein1,adhikari3,lewenstein2,adhikari2}. Initially a conjecture that SPA leads to separable states has been posed \cite{korbicz}. Later, this conjecture was disproved by taking various counterexamples\cite{kye,stormer,hansen}. On the experimental side, SPA has been exploited to realize quantum channels\cite{kye2,kye3,kye4}. The main idea of SPA tells us that how much proportion of white noise is mixed to a non-physical operator $\Lambda$ so that the approximated operator $\tilde{\Lambda}$ is completely positive. This make $\tilde{\Lambda}$ a physically realizable operator. The approximated operator $\tilde{\Lambda}$ can be written as\cite{horodecki3,bae},
	\begin{eqnarray}
	\tilde{\Lambda}=(1-p)\Lambda+pD,~~0\leq p\leq 1
	\end{eqnarray}
	where $D[\rho]=\frac{I_d}{d}$ is the depolarizing map and $I_d$ denote the $d$-dimensional identity matrix.
	In particular, partial transposition operation is a non-physical operation and thus, to make it a physically realizable operator we will perform SPA on it. Now, the SPA-PT map can be written as \cite{horodecki5},
	\begin{equation}
	\widetilde{id \otimes T}=p\frac{I\otimes I}{d^2}+(1-p)(I\otimes T)
	\end{equation}
	where p denotes the minimum value  for which $\widetilde{id \otimes T}$ becomes a positive semidefinite operator. When this completely positive operator $\widetilde{id \otimes T}$ acts on a density matrix $\rho^{in}$, it gives another density matrix $\rho^{out}$ i.e. $(id \otimes T)(\rho^{in})=\rho^{out}$.
	The eigenvalues of the output density matrix $\rho^{out}$ are important as the minimum eigenvalue of $\rho^{out}$ help us detecting entanglement physically. In general, a state $\rho$ in $d\otimes d$ dimensional system is separable if and only if the minimum eigenvalue of $\widetilde{id \otimes T}(\rho)$ is greater than or equals to $\frac{d^2\lambda}{d^4\lambda+1}$ where $-\lambda<0$ denotes the most negative eigenvalue when the induced map $[(I\otimes I)\otimes (I\otimes T)]$ acts on the maximally entangled state of the form, $\frac{1}{\sqrt{d^2}}\sum_{i=1}^{d^2}{|i\rangle |i\rangle}$.\\
	This paper is organized as follows: In Sec. II, we review two popular measure of entanglement: negtaivity and concurrence. In Sec. III, we define structured negativity as a physically realizable enatnglement measure and have shown that it satisfies the conditions of a valid entanglement monotone. In Sec. IV, we establish a relationship between negativity and structured negativity and give few examples to support our result. We conclude in Sec. V.

	\section{Measures of entanglement: Concurrence and Negativity}
	\noindent \textbf{Concurrence:} A very popular measure for the quantification of bipartite quantum correlations is the concurrence\cite{wootters,wootters1,wootters2}. For any two-qubit pure state $|\psi \rangle^{2\otimes 2}$, it is defined as,
	\begin{eqnarray}
	C(|\psi \rangle^{2\otimes 2})=\sqrt{2(1-Tr(\rho_{A}^{2}))}
	\end{eqnarray}
	where $\rho_{A}$ is the reduced state of $|\psi \rangle^{2\otimes 2}$. 
	Concurrence for the two-qubit mixed state may be defined as,
	\begin{eqnarray}
	C(\rho)=max(0,\sqrt{\lambda_{1}}-\sqrt{\lambda_{2}}-\sqrt{\lambda_{3}}-\sqrt{\lambda_{4}})
	\end{eqnarray}
	where	$\lambda_{i}^{s}$ are the eigenvalues of $\rho\tilde{\rho}$ and arranged in descending order. Here, $\tilde{\rho}$=$(\sigma_{y}\otimes\sigma_{y})\rho(\sigma_{y}\otimes\sigma_{y})$. Generalizing the definition of concurrence for $d_1\otimes d_2$ dimensional bipartite quantum system  \cite{rungta2}, we have
	\begin{eqnarray}
	C(|\psi\rangle^{d_1\otimes d_2})=\sqrt{2\nu_{d_1}\nu_{d_2}[1-Tr(\rho_{A}^{2})]}
	\end{eqnarray}
	where $\rho_{A}$ is the reduced state of $|\psi \rangle^{d_1\otimes d_2}$ and $\nu_{d_1}$ and $\nu_{d_2}$ are positive constants. Except for two-qubit system, we don't have any closed formula for bipartite $d_1\otimes d_2$ dimensional mixed state. For higher dimensional mixed states, we can estimate the amount of entanglement through the lower bound of the concurrence\cite{albevario,mintert,adhikari3}.
	
	\noindent \textbf{Negativity:} Negativity is another measure of entanglement based on the negative eigenvalues of the partially transposed matrix. It was first introduced as an entanglement measure by Vidal and Werner \cite{vidal}. The negativity for d-dimensional system described by the density operator $\rho$ may be defined as \cite{lee},
	\begin{eqnarray}
	N(\rho)=\frac{||\rho^{T_B}||_1-1}{d-1}=\frac{2}{d-1}\sum_{\lambda_i<0}{|\lambda_i(\rho^{T_B})|}
	\label{negativity}
	\end{eqnarray}
	where $||.||_1$ denotes trace norm and $\rho^{T_B}$ is the partial transposition of the density matrix $\rho$. The entanglement measure, negativity is useful in comparision to concurrence because unlike concurrence, it depends on the negative eigenvalues of the partial transposed state and thus exact value of negativity can be calculated very easily even for higher dimensional ntertxed quantum systems. Although negativity can be calculated exactly in theory for any arbitrary dimensional system but it cannot be implimented in laboratory. The reason behind this is that the partial transposition operation represent a positive but not a completely positive map.\\	
	To get rid of the difficulty of implimenting negativity in experiment, we have used another entanglement measure based on SPA-PT, defined in the next section.
	\section{Structured Negativity}
	\noindent Partial transposition map ($I\otimes T$) corresponds to a positive but not completely positive map and thus it cannot be implimented in the laboratory. To make PT map physically realizable, we use SPA-PT operation that may transform a positive but not completely positive map into a completely positive map. For $d\otimes d$ system described by the density operator $\rho$, the SPA-PT of the state $\rho$ denoted as $\tilde{\rho}$ and it may be expressed as \cite{horodecki5},
	\begin{equation}
	\tilde{\rho}=\frac{d}{d^3+1}I\otimes I +\frac{1}{d^3+1}[I\otimes T](\rho)
	\label{spapt}
	\end{equation}
	where, $I$ denote the identity matrix in $d\otimes d$ system.\\
	The state $\rho$ is separable if and only if \cite{horodecki5}
	\begin{eqnarray}
	\lambda_{min}(\tilde{\rho}) \geq \frac{d}{d^3+1}
	\label{separability}
	\end{eqnarray}
	where $\lambda_{min}(\tilde{\rho})$ denote the minimum eigenvalue of $\tilde{\rho}$. Otherwise the state $\rho$ is entangled.\\
	Now we are in a position to define a new measure of entanglement using the separability criteria given in (\ref{separability}). We may term this new measure of entanglement as structured negativity and it is denoted by $N_S({\rho})$. Therefore, for $d\otimes d$ system, the structured negativity may be defined as,
	\begin{eqnarray}
	N_S({\rho})=K.max\{\frac{d}{d^3+1}-\lambda_{min}(\tilde{\rho}),0\}
	\label{structured negativity}
	\end{eqnarray}
	where $K=d(d^3+1)$.\\
	To show $N_S({\rho})$, a valid measure of entanglement, we need to show that it satisfies few properties.
	\begin{enumerate}
		\item $N_S({\rho})$ vanishes if $\rho$ is separable.\\
		\textbf{Proof}: In $d\otimes d$ quantum system, if $\rho$ is separable then $\lambda_{min}(\tilde{\rho}) \geq \frac{d}{d^3+1}$\cite{horodecki5}. Thus, max$\{\frac{d}{d^3+1}-\lambda_{min}(\tilde{\rho}),0\}$=0. Hence, $N_S({\rho})=0$
		
		\item $N_S({\rho})$ is invariant under local unitary transformation.\\
		\textbf{Proof:}	$N_S({\rho})$ is invariant under a local change of basis since eigenvalues does not change under local change of basis \cite{ziman,anuma}. Thus, $N_S({\rho})=N_S(U_A \otimes U_B {\rho} U_A^{\dagger} \otimes U_B^{\dagger})$ where $U_A$ and $U_B$ denotes the unitaries acting on the subsystem A and B respectively.
		
		\item $N_S({\rho})$ satisfies convexity property i.e.
		\begin{eqnarray}
		N_S(\sum_k{p_k{\rho_k}}) \leq \sum_k{p_k}N_S({\rho_k})
		\end{eqnarray}
		\textbf{Proof:} Let us consider a quantum state described by the density operator $\rho=\sum_k{p_k{\rho_k}}$, $(0\leq p_k \leq 1)$. The structured negativity of $\rho$ is given by
		\begin{eqnarray}
		N_S(\rho)=N_S(\sum_k{p_k{\rho_k}})=K[\frac{d}{d^3+1}-\lambda_{min}(\tilde{\rho})]
		\label{convexity}
		\end{eqnarray}
		Using Lemma 1, RHS of equation (\ref{convexity}) may be re-expressed as
		\begin{eqnarray}
		N_S(\sum_k{p_k{\rho_k}})=K[\frac{d}{d^3+1}-\lambda_{min}(p_1\tilde{\rho_1}+\sum_{k \neq 1}{p_k\tilde{\rho_k}})]
		\label{weyl}
		\end{eqnarray}
		Using Weyl's inequality in (\ref{weyl}), we get
		\begin{eqnarray}
		N_S(\sum_k{p_k{\rho_k}}) \leq K [\frac{d}{d^3+1}-(\lambda_{min}(p_1\tilde{\rho_1})+\lambda_{min}(\sum_{k \neq 1}{p_k\tilde{\rho_k}})))]
		\label{eq14}
		\end{eqnarray}
		Using Weyl's inequality repeatedly (k-1) times, equation (\ref{eq14}) may be re-written as,
		\begin{eqnarray}
		N_S(\sum_k{p_k{\rho_k}}) &\leq& K [\frac{d}{d^3+1}-\sum_{k}\lambda_{min}({p_k\tilde{\rho_k}}))]\nonumber\\
		&=& K[\sum_{k}{p_k}(\frac{d}{d^3+1}-\lambda_{min}({\tilde{\rho_k}}))]\nonumber\\
		&=& \sum_k{p_k}.K.(\frac{d}{d^3+1}-\lambda_{min}({\tilde{\rho_k}}))\nonumber\\
		&=& \sum_k{p_k}{N_S({\rho_k})}
		\end{eqnarray}
		\item $N_S({\rho})$ does not increase on average under LOCC\cite{plenio1} i.e.
		\begin{eqnarray}
		N_S(\rho) \geq \sum_i{p_i}N_S[(K_i\otimes I)\rho (K_i^{\dagger}\otimes I)]
		\label{locc}
		\end{eqnarray}
		where $K_i$ are the Kraus operators.\\
		\textbf{Proof:} Let us consider an entangled state described by a density operator $\rho$. Now, consider the right hand side of (\ref{locc}) that can be expressed as
		\begin{eqnarray}
		&&\sum_i{p_i}N_S[(K_i\otimes I)\rho (K_i^{\dagger}\otimes I)]\nonumber \\&=&K\sum_ip_i[\frac{d}{d^3+1}-\lambda_{min}(\widetilde{(K_i\otimes I)\rho (K_i^{\dagger}\otimes I))}]\nonumber\\&=&K[\frac{d}{d^3+1}-\sum_i{p_i\lambda_{min}(\widetilde{(K_i\otimes I)\rho (K_i^{\dagger}\otimes I))})}]\nonumber \\&\geq& K[\frac{d}{d^3+1}-\sum_i{\lambda_{min}(\widetilde{(K_i\otimes I)\rho (K_i^{\dagger}\otimes I))})}]
		\label{eq1}
		\end{eqnarray}
		The first step follows from the definition of structured negativity given in (\ref{structured negativity}). In the third step, the inequality follows from the fact $0\leq p_i\leq 1$.\\ Using Weyl's inequality, (\ref{eq1}) may be expressed as
		\begin{eqnarray}
			&&\sum_i{p_i}N_S[(K_i\otimes I)\rho (K_i^{\dagger}\otimes I)]\nonumber \\&\geq&K[\frac{d}{d^3+1}-\lambda_{min}[\sum_i{(\widetilde{(K_i\otimes I)\rho (K_i^{\dagger}\otimes I))}}]\nonumber\\&=& K[\frac{d}{d^3+1}-\lambda_{min}[\sum_i(\frac{d}{d^3+1}I\otimes I\nonumber\\&+&\frac{1}{d^3+1}(I\otimes T)((K_i\otimes I)\rho(K_i^{\dagger}\otimes I)))]]
			\label{eq17}
			\end{eqnarray}
			In (\ref{eq17}), the second step follows from the definition (\ref{spapt}) of SPA-PT  of $(K_i\otimes I)\rho(K_i^{\dagger}\otimes I)$.
			Let us assume that the entangled state described by the density operator $\rho$ may be evolved as $\rho^{'}=\sum_{i=1}^{m}((K_i\otimes I)\rho (K_i^{\dagger}\otimes I))$, where $m$ denote the number of Kraus operators.\\
			Then, (\ref{eq17}) can be re-expressed as			\begin{eqnarray}
			&&\sum_i{p_i}N_S[(K_i\otimes I)\rho (K_i^{\dagger}\otimes I)]\nonumber\\&\geq&K[\frac{d}{d^3+1}-\lambda_{min}[\frac{md}{d^3+1}I\otimes I+\frac{1}{d^3+1}(I\otimes T)\rho^{'}]]\nonumber\\&=&K[\frac{d}{d^3+1}-\lambda_{min}[\frac{md}{d^3+1}I\otimes I+\frac{1}{d^3+1}({\rho^{'}}^{T_B})]]
			\label{eq2}
			\nonumber \\&\geq&K[\frac{d}{d^3+1}-\lambda_{min}[(\tilde{\rho^{'}})+\frac{(m-1)d}{d^3+1}I\otimes I] ]
	\label{eq3}
	\end{eqnarray}
	Using the upper bound of Weyl's inequality, (\ref{eq3}) may be expressed as
	\begin{eqnarray}
	&&\sum_i{p_i}N_S[(K_i\otimes I)\rho (K_i^{\dagger}\otimes I)]\nonumber \\&\geq&K[\frac{d}{d^3+1}-\lambda_{min}(\tilde{\rho^{'}})-\frac{(m-1)d}{d^3+1}]
	\label{eq4}
	\end{eqnarray}
	To show	$N_S(\rho) \geq \sum_i{p_i}N_S[(K_i\otimes I)\rho (K_i^{\dagger}\otimes I)]$, it is sufficient to show that $N_S(\rho)-\sum_i{p_i}N_S[(K_i\otimes I)\rho (K_i^{\dagger}\otimes I)]\geq0$.\\
	Using (\ref{structured negativity}) and(\ref{eq4}), we have
	\begin{eqnarray}
	&&N_S(\rho)-\sum_i{p_i}N_S[(K_i\otimes I)\rho (K_i^{\dagger}\otimes I)]\nonumber\\&\geq& K[\lambda_{min}(\tilde{\rho^{'}})+\frac{(m-1)d}{d^3+1}-\lambda_{min}(\tilde{\rho})]
	\label{eq5}
	\end{eqnarray}
	For an entangled state $\rho$, we have
	\begin{eqnarray}
	\lambda_{min}(\tilde{\rho})<\frac{d}{d^3+1}
	\label{entangled}
	\end{eqnarray}
	where $\tilde{\rho}$ denote the SPA-PT of $\rho$.\\
	Using (\ref{entangled}), the inequality (\ref{eq5}) reduces to
	\begin{eqnarray}
	&&N_S(\rho)-\sum_i{p_i}N_S[(K_i\otimes I)\rho (K_i^{\dagger}\otimes I)]\nonumber\\&\geq& K[\lambda_{min}(\tilde{\rho^{'}})+\frac{(m-2)d}{d^3+1}]
	\end{eqnarray}
Since $m\geq 2$ so, $\lambda_{min}(\tilde{\rho^{'}})+\frac{(m-2)d}{d^3+1} \geq 0$. Thus
\begin{eqnarray}
&&N_S(\rho)-\sum_i{p_i}N_S[(K_i\otimes I)\rho (K_i^{\dagger}\otimes I)]\geq 0\nonumber\\&\implies&N_S(\rho)\geq \sum_i{p_i}N_S[(K_i\otimes I)\rho (K_i^{\dagger}\otimes I)]
\end{eqnarray}
Hence proved.
\end{enumerate}
	
	\section{Relation between negativity and structured negativity}
	In this section, we derive the relationship between the negativity and structured negativity of a given quantum state $\rho$.\\
	\textbf{Result-1:} For any quantum state $\rho$ in $d\otimes d$ system, the relation between negativity and structured negativity is given by
	\begin{eqnarray}
	N(\rho)\leq 2(1-\frac{1}{d})N_S({\rho})
	\label{finalnrho}
	\end{eqnarray}
	\textbf{Proof:}  For any two Hermitian matrices $A,B \in M_n$ , Weyl's inequality may be defined as \cite{weyl,adhikari5}
	\begin{eqnarray}
	\lambda_k(A+B)\leq \lambda_{k+j}(A)+\lambda_{n-j}(B)
	\label{eq18}
	\end{eqnarray}
	where eigenvalues of the matrices $A$, $B$ and $A+B$ are arranged in increasing order.\\
	For k=1, Weyl's inequality reduces to
	\begin{eqnarray}
	\lambda_{min}(A+B)\leq \lambda_{1+j}(A)+\lambda_{n-j}(B)
	\end{eqnarray}
	for $j=0,1...n-1$.\\
	If $\tilde{\rho}$ denote the SPA-PT of $\rho \in M_{d^2}$, then by taking $A=\frac{1}{d^3+1}({I\otimes T)}\rho \equiv\frac{1}{d^3+1}\rho^{T_B}$ and $B=\frac{d}{d^3+1}I \otimes I$, (\ref{eq18}) reduces to
	\begin{eqnarray}
	\lambda_{min}({\tilde{\rho}})\leq \lambda_{1+j}(\frac{1}{d^3+1}\rho^{T_B})+\frac{d}{d^3+1}\lambda_{d^2-j}(I\otimes I)
	\label{k1}
	\end{eqnarray}
	for $j=0,1...n-1$.\\
	Let us suppose that $(I\otimes T)\rho$ has $q (\leq (d-1)^2)$ number of negative eigenvalues, if $\rho$ is an entangled state. \\
	Putting $j=0,1...q-1$ in (\ref{k1}) and adding, we get
	\begin{eqnarray}
	q\lambda_{min}(\tilde{\rho}) \leq \frac{1}{d^3+1}\sum_{i=1,\lambda_i <0}^{q}{\lambda_i(\rho^{T_B})}+\frac{dq}{d^3+1}
	\label{eq20}
	\end{eqnarray}
	Exploiting the definition of negativity given in (\ref{negativity}), the above equation (\ref{eq20}) reduces to
	\begin{eqnarray}
	\lambda_{min}(\tilde{\rho}) \leq -\frac{(d-1)N(\rho)}{2q(d^3+1)}+\frac{d}{d^3+1}
	\label{nrho}
	\end{eqnarray}
	If $\rho$ is an entangled state, then the minimum eigenvalue of $\tilde{\rho}$ (given in (\ref{structured negativity})) may be expressed as,
	\begin{eqnarray}
	\lambda_{min}({\tilde{\rho}})=\frac{d^2-N_S({{\rho}})}{d(d^3+1)}
	\label{n1rho}
	\end{eqnarray}
	Substituting value of $\lambda_{min}({\tilde{\rho}})$ in (\ref{nrho}), we get
	\begin{eqnarray}
	\frac{d^2-N_S({{\rho}})}{d(d^3+1)} \leq -\frac{(d-1)N(\rho)}{2q(d^3+1)}+\frac{d}{d^3+1}
	\end{eqnarray}
	After simplification, we get the required relation,
	\begin{eqnarray}
	N_S({\rho}) \geq \frac{d(d-1)}{2q}N(\rho)
	\label{nrhon1rho}
	\end{eqnarray}
	Thus, the structured negativity of a given quantum state is always greater than or equals to $\frac{d(d-1)}{2q}$ times to it's negativity.
	The number of negative eigenvalues of $\rho^{T_B}$ for a $d\times d$ dimensional system are atmost $(d-1)^2$. Substituting $q\leq (d-1)^2$, (\ref{nrhon1rho}) becomes,
	\begin{eqnarray}
	N_S({\rho})\geq \frac{d}{2(d-1)}N(\rho)
	\end{eqnarray}
	This implies,
	\begin{eqnarray}
	N(\rho)\leq 2(1-\frac{1}{d})N_S({\rho})
	\label{finalnrho}
	\end{eqnarray}
	Hence proved.\\
	\textbf{Remark:} For sufficiently large d, we have $1-\frac{1}{d} \approx 1$. Thus for higher dimensional system, the inequality (\ref{finalnrho}) reduces to $N(\rho) \leq 2N_S({\rho})$.
	\subsection{Examples}
	\noindent As we have mentioned in Sec.II, that there does not exist any closed formula for concurrence of arbitrary dimensional bipartite mixed state but neverthless, there exist lower bound of the concurrence for arbitrary $d\otimes d$ dimensional system \cite{mintert,albevario,adhikari3}. One such lower bound of concurrence obtained in \cite{albevario} which may be given by
	\begin{eqnarray}
	C(\rho)&\geq& \sqrt{\frac{2}{d(d-1)}}(max(||\rho^{T_B}||,||R(\rho)||)-1)\nonumber\\
	&\equiv& C_{lb}(\rho)
	\end{eqnarray}
	where $C(\rho)$ and $C_{lb}$ respectively denote the concurrence and lower bound of the concurrence of the state $\rho$ in arbitrary $d\otimes d$ dimensional system. $||.||$ denote the trace norm of $(.)$. $R(\rho)$ and $\rho^{T_B}$ respectively denote the realignment operation and partial transposition operation with respect to subsystem $B$.\\
	In this subsection, we will provide few examples by which we can show that Result-1 is indeed true. For the given state $\rho$, we have compared three measures of entanglement such as negativity $(N(\rho))$, structured negativity $(N_S(\rho))$ and the lower bound of the concurrence $(C_{lb}(\rho))$ and found that structured negativity is always greater than or equals negativity. In few cases structured negativity is greater than the lower bound of concurrence. This indicates that the new measure of entanglement quantifies the amount of entanglement better than negativity and for some cases, better than estimator of entanglement $C_{lb}$. Also, we have observed that for $d\otimes d$ dimensional system, negativity and structured negativity coincides when the number of negative eigenvalues of the partially transposed matrix are $\frac{d(d-1)}{2}$ i.e equality holds in (\ref{nrhon1rho}), when $q=\frac{d(d-1)}{2}$, and strict inequality holds when $q\neq \frac{d(d-1)}{2}$.\\

	\subsubsection{Two-qubit Werner state}
	Let us consider a two qubit Werner state defined as \cite{adhikari5,horodecki9}
	\begin{eqnarray}
	\rho_W=F|\psi^{-}\rangle \langle \psi^{-}|+(1-F)\frac{I}{4}
	\end{eqnarray}
	where
	\begin{eqnarray}
	|\psi^{-}\rangle=\frac{1}{\sqrt{2}}|01\rangle -|10\rangle
	\end{eqnarray}
	The family of Werner states are the only states invariant under the transformation \cite{horodecki9}
	\begin{eqnarray}
	\rho_W \longrightarrow U\otimes U \rho_W U^{\dagger} \otimes U^{\dagger}
	\end{eqnarray}
	where $U$ is a unitary transformation.
	The state is entangled for $\frac{1}{3} <F\leq1$. Negativity of $\rho_W$ is $\frac{3F-1}{2}$  and structured negativity, $N_S(\rho_W)=18[\frac{2}{9}+\frac{1}{12}(-3+F)]$. From Fig-1, Result-1 is verified.
	\begin{figure}[h]
		\centering
		\includegraphics[scale=0.35]{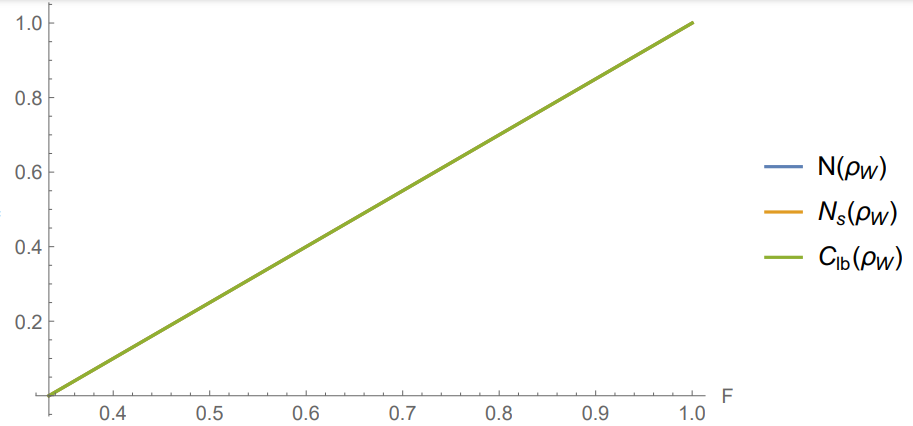}
		\caption{The negativity, structured negativity and lower bound of concurrence for two-qubit Werner state coincide with each other. Therefore, the equality in (\ref{finalnrho}) has been achieved for a family of Werner state.}
	\end{figure}
	
	\subsubsection{Two-qubit MEMS state}
	Consider a maximally entangled mixed state (MEMS) introduced by Munro et.al. \cite{munro,adhikari6}
	\begin{eqnarray}
	\rho_{MEMS}=
	\begin{pmatrix}
	h(C) & 0 & 0 & \frac{C}{2}\\
	0 & 1-2h(C) & 0 & 0\\
	0 & 0 & 0 & 0\\
	\frac{C}{2} & 0 & 0 & h(C)
	\end{pmatrix}
	\end{eqnarray}
	where,
	\begin{eqnarray}
	h(C)=
	\begin{cases}
	\frac{C}{2} & C\geq \frac{2}{3}\\
	\frac{1}{3} & C<\frac{2}{3}
	\end{cases}
	\end{eqnarray}	
where C denotes the concurrence of $\rho_{MEMS}$.\\
For $C\geq \frac{2}{3}$, $N(\rho_{MEMS})=-1+C+\sqrt{1-2C+2C^2}$,  $N_S(\rho_{MEMS})=18(\frac{2}{9}+\frac{1}{18}(-5+C+\sqrt{1-2C+2C^2}))$. From Fig-2, it can be seen that negativity, structured negativity and $C_{lb}(\rho_{MEMS})$ of the state $\rho_{MEMS}$ coincide for $C\geq \frac{2}{3}$.
	\begin{figure}[h]
		\centering
		\includegraphics[scale=0.35]{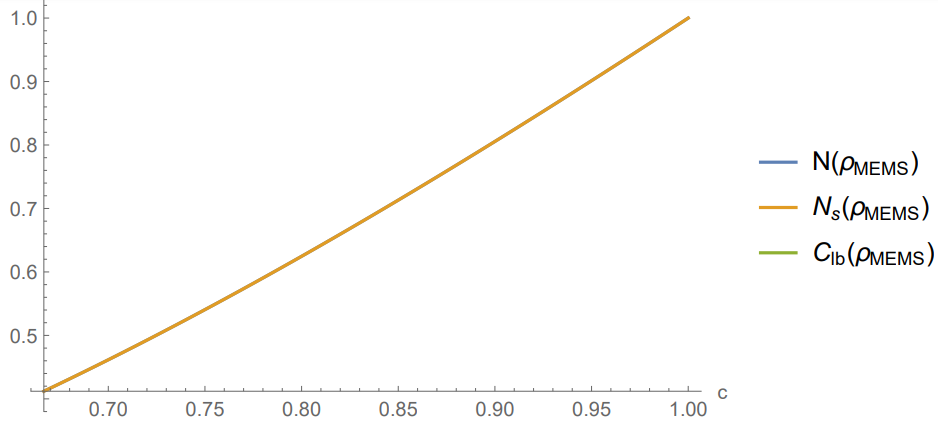}
		\caption{The negativity, structured negativity and lower bound of concurrence concides for the state  $\rho_{MEMS}$ for $C\geq\frac{2}{3}$.}
		\label{MEMS1}
	\end{figure}
	Also for $C<\frac{2}{3}$ negativity and structured negativity of $\rho_{MEMS}$ are given by: $N(\rho_{MEMS})=\frac{1}{3}(-1+\sqrt{1+9C^2})$ and $N_S(\rho_{MEMS})=18(\frac{2}{9}+\frac{1}{54}(-13+\sqrt{1+9C^2}))$. From Fig-3, it can be seen that negativity is same as the structured negativity for $\rho_{MEMS}$($C<\frac{2}{3}$).
	\begin{figure}[h]
		\centering
		\includegraphics[scale=0.35]{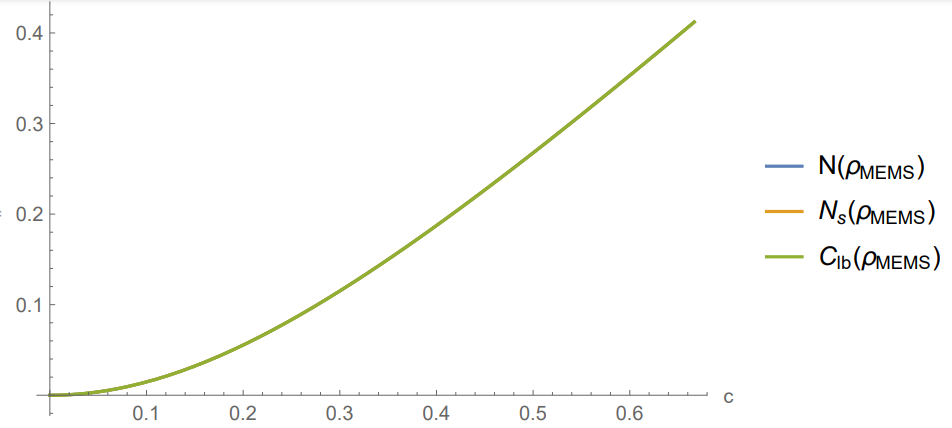}
		\caption{The negativity, structured negativity and lower bound of concurrence concides for the state  $\rho_{MEMS}$ for $C< \frac{2}{3}$.}
		\label{MEMS1}
	\end{figure}

	\subsubsection{Two-qutrit state described by the density operator $\rho_{a}$}
	Consider a two-qutrit state defined in \cite{swapan}, which is described by the density operator
	\begin{eqnarray}
	\rho_a=\frac{1}{5+2a^2}\sum_{i=1}^{3}{|\psi_i\rangle \langle \psi_i|},~~\frac{1}{\sqrt{2}}\leq a \leq 1
	\end{eqnarray}
	where, $|\psi_i\rangle=|0i\rangle-a|i0\rangle$, for $i=\{1,2\}$ and\\ $|\psi_3\rangle=\sum_{i=0}^{n-1}{|ii\rangle}$. \\
	For the state $\rho_a$ the negativity $N(\rho_a)$ and the structured negativity $N_S({\rho_a})$ can be calculated as
	\begin{eqnarray}
	N(\rho_a)&=&\frac{1}{5+2a^2}-\frac{1-2\sqrt{a}}{5+2a^2}\nonumber\\&-&\frac{1+a^2-\sqrt{5-2a^2+a^4}}{5+2a^2}\\
	N_S({\rho_a})&=&84[\frac{3}{28}-\frac{7+3a^2}{14(5+2a^2)}]
	\end{eqnarray} The comparision between $N(\rho_a)$, $N_S(\rho_a)$, $C_{lb}(\rho_{a})$ for the two-qutrit state $\rho_a$ has been studied in Fig-4.
	\begin{figure}[h]
		\centering
		\includegraphics[scale=0.25]{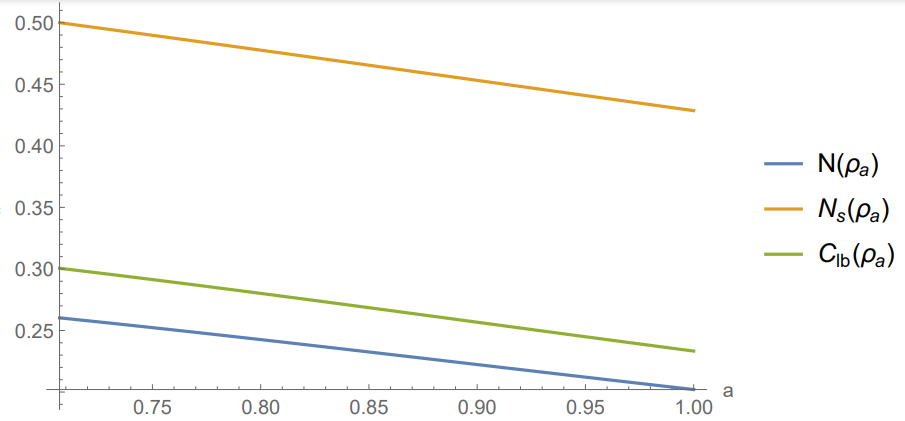}
		\caption{The structured negativity of the state $\rho_a$ is greater than the negativity as well as the lower bound of concurrence.}
		\label{MEMS2}
	\end{figure}
	
		\subsubsection{Two-qutrit $\alpha$ state}
Consider a two-qutrit state defined by\cite{horodecki6},
\begin{eqnarray}
\rho_{\alpha}=\frac{2}{7}|\psi^{+}\rangle \langle \psi^{+}|+\frac{\alpha}{7}\sigma_{+}+\frac{5-\alpha}{7}{\sigma_{-}}c, ~~2\leq \alpha\leq 5
\end{eqnarray}
where,
\begin{eqnarray}
|\psi^{+}\rangle&=&\frac{1}{\sqrt{3}}[|00\rangle+|11\rangle+|22\rangle]\nonumber\\
\sigma_{+}&=&\frac{1}{3}(|01\rangle \langle 01|+|12\rangle \langle 12|+|20\rangle \langle 20|)\nonumber\\
\sigma_{-}&=&\frac{1}{3}(|10\rangle \langle 10|+|21\rangle \langle 21|+|02\rangle \langle 02|)
\end{eqnarray}
The given state $\rho_{\alpha}$ is NPTES for $4\leq \alpha \leq5$.\\
For the state $\rho_{\alpha}$, 
In this example, first consider $\rho_{\alpha}$ for $4\leq \alpha \leq 5$. The SPA-PT of $\rho_{\alpha}$ is given by, $N(\rho_a)$ and $N_S({\rho_a})$ may be calculated as
\begin{eqnarray}
	N(\rho_{\alpha})&=&-\frac{1}{14}[-5+\sqrt{41-20\alpha+4\alpha^{2}}]\\
	N_S({\rho_{\alpha}})&=&84[\frac{3}{28}-\frac{-131+\sqrt{41-20\alpha+4\alpha^{2}}}{1176}]
\end{eqnarray}
The comparision between $N(\rho_{\alpha})$, $N_S(\rho_{\alpha})$ and $C_{lb}(\rho_{\alpha})$ for the  state $\rho_{\alpha}$ has been studied in fig-5.
\begin{figure}[h]
	\centering
	\includegraphics[scale=0.25]{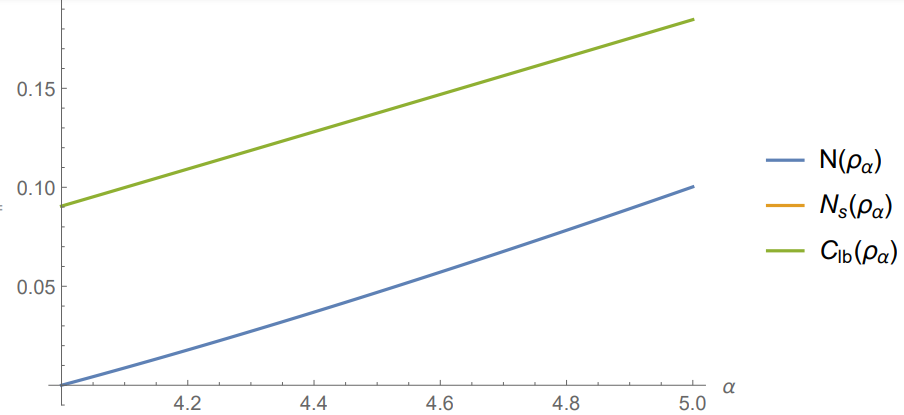}
	\caption{The negativity and structured negativity coincide for the state $\rho_{\alpha}$. In this case, $C_{lb}(\rho_{\alpha})$ is greater than the structured negativity. }
	\label{ex4}
\end{figure}

	\section{Conclusion}
	\noindent To summarize, we have defined an entanglement measure based on the minimum eigenvalue of the SPA-PT of the arbitrary dimensional bipartite quantum system. We call the introduced measure as structured negativity and proved that it satisfies the properties of a valid entanglement measure. Since, SPA-PT is a completely positive map, so, the proposed measure of entanglement can be realized in an experiment. The introduced measure of entanglement provides an advantage over negativity. This is due to the fact that negativity depends on the sum of the negative eigenvalues of the non-physical partial transposition operation whereas structured negativity depends on the minimum eigenvalue of SPA of the partially transposed matrix. We have established a relation between negativity and structured negativity and found that the negativity and structured negativity coincides for large number of two-qubit systems. Thus, we conjecture that the negativity and structured negativity coincides when $q=\frac{d(d-1)}{2}$. In the end, we have shown that the structured negativity is always greater than or equals to the negativity. In few cases, we found that the structured negativity gives better result than lower bound of concurrence $C_{lb}$. Thus, it is a better measure of entanglement than negativity always but in few cases, it will be better than lower bound of concurrence $C_{lb}$.

	
	\section{Acknowledgement}
	A.K. would like to acknowledge the financial support from CSIR. This work is supported by CSIR File No. 08/133(0027)/2018-EMR-1.
	
	\section{Appendix-1}
	\textbf{Lemma 1:} Any $d\otimes d$ system which is described by the density operator $\rho=\sum_k{p_k\rho_k}$, it's SPA-PT is given by,
	\begin{eqnarray}
	\tilde{\rho}=\sum_k{p_k\tilde{\rho_k}}
	\end{eqnarray}
	\textbf{Proof:} Consider a bipartite state $\rho=\sum_k{p_k\rho_k}$ in $d\otimes d$ system, then it's SPA-PT may be written as
	\begin{eqnarray}
	\tilde{\rho}&=&\frac{d}{d^3+1}(I\otimes I)+\frac{1}{d^3+1}(I\otimes T)\rho\nonumber\\
	&=&\frac{d}{d^3+1}(I\otimes I)+\frac{1}{d^3+1}(I\otimes T)\sum_k{p_k\rho_k}\nonumber\\
	&=&\frac{d}{d^3+1}(I\otimes I)+\sum_{k}{p_k}\frac{1}{d^3+1}(I\otimes T)\rho_k\nonumber\\
	&=&\sum_{k}{p_k\frac{d}{d^3+1}}(I\otimes I)+\sum_{k}{p_k}\frac{1}{d^3+1}(I\otimes T)\rho_k\nonumber\\
	&=&\sum_{k}{p_k[}\frac{d}{d^3+1}(I\otimes I)+\frac{1}{d^3+1}(I\otimes T)\rho_k]\nonumber\\
	&=& \sum_{k}{p_k\tilde{\rho_k}}
	\end{eqnarray}



	

\end{document}